\documentclass[12pt,preprint]{aastex}

\def\etal{{\it et al. }}

\def\gtsim{\ {\raise-0.5ex\hbox{$\buildrel>\over\sim$}}\ }
\def\ltsim{\ {\raise-0.5ex\hbox{$\buildrel<\over\sim$}}\ }

\begin{document}

\title{
Millimagnitude Photometry for Transiting Extrasolar Planetary Candidates IV: 
The Puzzle of the Extremely Red OGLE-TR-82 Primary Solved\altaffilmark{11}}

\author{Sergio Hoyer\altaffilmark{2,1},
Sebasti\'an Ram\'{\i}rez Alegr\'{\i}a\altaffilmark{1},
Valentin D. Ivanov\altaffilmark{3},
Dante Minniti \altaffilmark{1},
Grzegorz Pietrzynski\altaffilmark{4,5},
Mar\'ia Teresa Ru\'iz\altaffilmark{2},
Wolfgang Gieren\altaffilmark{4},
Andrzej Udalski\altaffilmark{5},
Manuela Zoccali\altaffilmark{1},
Rodrigo Carrasco\altaffilmark{6},
Rodrigo F. D\'{\i}az\altaffilmark{7},
Jos\'e Miguel Fern\'andez\altaffilmark{1,8},
Jos\'e Gallardo\altaffilmark{9},
Marina Rejkuba\altaffilmark{10},
Felipe P\'erez\altaffilmark{1}
}

\altaffiltext{1}{Department of Astronomy, Pontificia Universidad Cat\'olica, 
Casilla 306, Santiago 22, Chile; sramirez@astro.puc.cl, dante@astro.puc.cl, mzoccali@astro.puc.cl, fperez@astro.puc.cl}
\altaffiltext{2}{Department of Astronomy, Universidad de Chile, Casilla 36-D, Santiago, Chile; shoyer@das.uchile.cl, mtruiz@das.uchile.cl}
\altaffiltext{3}{European Southern Observatory, , Alonso de Cordova 3107, Vitacura,
Casilla 19001, Santiago 19, Chile; vivanov@eso.org}
\altaffiltext{4}{Department of Physics, Universidad de Concepci\'on, Casilla 160-C, Concepci\'on, Chile;  pietrzyn@hubble.cfm.udec.cl, wgieren@astro-udec.cl}
\altaffiltext{5}{Warsaw University Observatory, Al. Ujazdowskie 4, 00-478 Waszawa, Poland; udalski@astrouw.edu.pl}
\altaffiltext{6}{Gemini Observatory, Southern Operations Center, AURA, Casilla 603, La Serena, Chile, rcarrasco@gemini.edu}
\altaffiltext{7}{Instituto de Astronom\'{\i}a y F\'{\i}sica del Espacio, CONICET- UBA, CC 67 - Suc 28, Buenos Aires C1428ZAA, Argentina; rodrigo@iafe.uba.ar}
\altaffiltext{8}{Harvard-Smithsonian Center for Astrophysics, 60 Garden St. Cambridge, MA 02138, USA; jfernand@gmail.com}
\altaffiltext{9}{Centre de Recherche Astronomique de Lyon, Ecole normale
superieure de Lyon, 46 allee de Italie, 69364 Lyon, France; jose.gallardo@ens-lyon.fr}
\altaffiltext{10}{European Southern Observatory, Karl-Schwarzschild-Str.2, 85748 Garching bei Muenchen, Germany; vivanov@eso.org, mrejkuba@eso.org}
\altaffiltext{11}{Based on observations collected with the 
GEMINI-S Telescope (queue observing, program GS-2005B-Q-9), with the
Very Large Telescope at Paranal Observatory 
(JMF and DM visiting observers), with the
ESO New Technology Telescope at La Silla Observatory (SR, FP, and DM visiting
observers) for the ESO Programmes 075.C-0427, 075.B-0414, and 076.C-0122.}

\begin{abstract}

We present precise new $V$, $I$, and $K$-band photometry for the
planetary transit candidate star OGLE-TR-82. Good seeing $V$-band
images acquired with VIMOS instrument at ESO Very Large Telescope (VLT) allowed us to measure $V=20.6\pm 0.03$~mag star in spite of the presence of a brighter neighbour about 1" away. This faint
magnitude answers the question why it has not been possible to 
measure radial velocities for this object.  

One transit of this star has been observed with GMOS-S instrument of
GEMINI-South telescope in $i$ and $g$-bands. The measurement of the
transit allows us to verify that this is not a false positive, to
confirm the transit amplitude measured by OGLE, and to improve the
ephemeris. The transit is well defined in $i$-band light curve, with a
depth of $A_i=0.034$~mag. It is however, less well defined, but deeper
($A_g = 0.1$~mag) in the $g$-band, in which the star is significantly
fainter.

The near-infrared photometry obtained with SofI array at the ESO New Technology Telescope (NTT)
yields $K=12.2\pm 0.1$, and $V-K=8.4\pm 0.1$, so red that it is unlike any other transit candidate 
studied before.  Due to the extreme nature of this object, we have not yet been able to measure velocities for this star, but based on 
the new data we consider two different possible configurations:
(1) a nearby M7V star, or (2) a blend with a very reddened distant red giant. 
The nearby M7V dwarf hypothesis would give a radius for the companion of $R_p=0.3\pm 0.1 ~R_J$, i.e. the size of Neptune.

Quantitative analysis of near-IR spectroscopy finally shows that OGLE-TR-82 is a distant, reddened metal poor early K giant.   This result is confirmed by direct comparison with stellar templates
that gives the best match for a K3III star.

Therefore, we discard the planetary nature of the companion. Based on all the new data, we conclude that this system is a main-sequence binary blended with a background
red giant.  As a case study, this kind of system that can mimic a planetary transit is a lesson to learn for future transit surveys.

\end{abstract}


\section{Introduction}

Because low mass stars are the most numerous stars in our Galaxy, in
spite of their dimness they 
provide an exciting possibility to detect transits of planets much smaller
than Jupiter. This has been pointed out by many authors
(see for example Pepper \& Gaudi 2005).  
The known planets in the Solar neighborhood show a very steep function of mass:
low mass planets are much more common around Solar type stars
than high mass planets (Butler \etal 2006).
Even though low mass planets are more difficult to detect than Jupiters,
there have been a few recent discoveries of ``Neptunes" or ``Super Earths" around
low mass stars (McArthur \etal 2004, Butler \etal 2004, 
Santos \etal 2004, Bonfils \etal 2006, Lovis \etal 2006, Beaulieu \etal 2006, Gould \etal 2006). 


More recently, 
Sahu \etal (2006) found a number of planetary candidates around low-mass
stars in the bulge fields.  The search for planets is being extended into
a new realm: M-dwarfs are the most numerous stars in the Galaxy,
and the transiting planets around such stars open a wealth
of new possibilities for ground and space-based planet searches.
The transit candidate OGLE-TR-82 (Udalski \etal 2002)
appears to be an extreme case, that is important to study in the context
of the future transit discoveries made by space missions like KEPLER and COROT.
We note also that there has been progress on the models triggered by these observational
discoveries as well. For example, the formation of Neptunes and super-Earths
has been recently studied by  
Baraffe \etal (2005, 2006), Kennedy \etal (2006) and Alibert \etal (2006).

\begin{figure}
\plotone{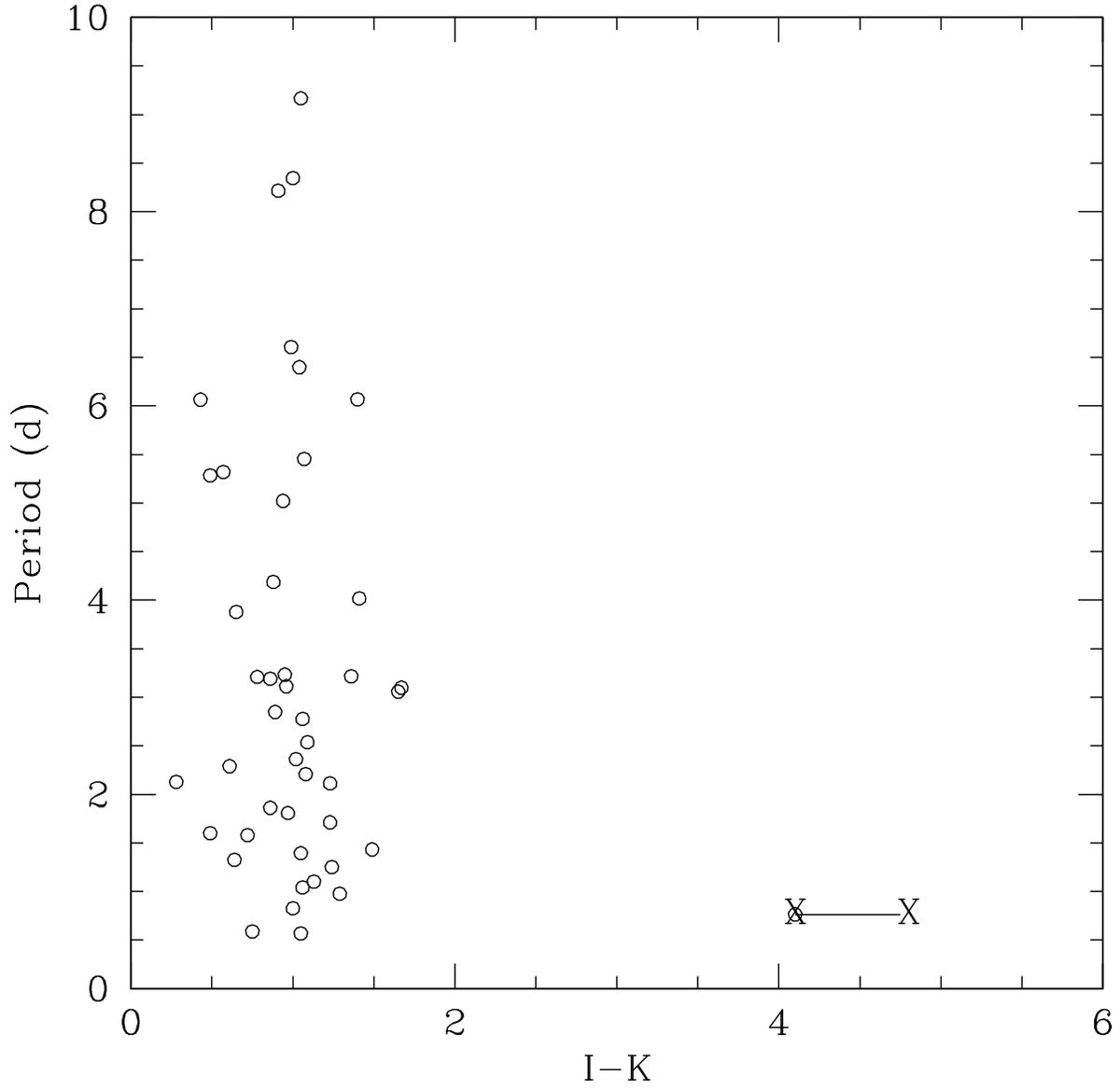}
\caption{
Period $vs$ $I-K$ color for OGLE transit candidates in the Carina region.
The reddest object is OGLE-TR-82, marked with the cross. The left cross on 
top of the circle is our $I-K$, and the right cross is from 2MASS.
}\label{uno}
\end{figure}

The OGLE search has provided the largest list of transiting candidates
(Udalski \etal 2002a, 2002b, 2002c, 2003). In particular,
Udalski \etal (2002c) discovered very low amplitude transits in the
$I=16.30$ magnitude  star OGLE-TR-82, located in the Carina region
of the Milky Way disk, at $RA(2000)=10:58:03.07$, $DEC(2000)=-61:34:25.8$. 
They monitored 22 individual transits on this star, measuring an amplitude 
$A_I=0.034$ mag, and a period $P=0.76416$ days.
They listed $R_p = 1.09~R_J$, making this a prime Hot Jupiter candidate.
Recently, Silve \& Cruz (2006) obtained $M_s=0.65 ~M_{\odot}$, and
 $R_p = 1.10~R_J$, also singling OGLE-TR-82-b as a possible planetary companion.
We followed-up OGLE candidates with infrared and optical photometry
with the aim of identifying and characterizing planetary companions
to low-mass stars (Gallardo \etal 2005, Fern\'andez \etal 2006, D\'{\i}az \etal 2006).
M-dwarfs with hot Jupiters should stand out in a period $vs$ color diagram
as objects with short periods and very red colors.
Early in the program it was recognized that OGLE-TR-82 was a very interesting
candidate. For example, Figure \ref{uno} shows the OGLE periods $vs$ the $I-K$
color for several candidates observed in the Carina fields.
OGLE-TR-82 is clearly unique in this sample: it is a very red and faint object, 
both according to 2MASS and to our own photometry.
The colors were consistent with a very late M-dwarf,
for which the low amplitude transits would imply a very small companion.
We then intensively followed up this candidate.
However, acquiring the complementary data to characterize OGLE-TR-82
and its companion has proven difficult due to the extreme nature of this object.

Pont \etal (2004) observed this candidate with UVES/FLAMES at the VLT,
arguing that this was possibly a K7-M0V star.
They obtained precise velocity measurements for several OGLE targets,
discovering three planets in their sample,
but they could find no CCF signal for OGLE-TR-82, concluding that this
was an unsolved case.  
Why is there no signal for this object? In this paper we answer the question
using optical and infrared photometric data acquired at GEMINI, VLT and NTT.

The future space missions like COROT and KEPLER would discover many faint transit
candidates, which would need to be followed up and confirmed. The most fundamental
follow-up observation is echelle spectroscopy in order to measure radial velocities
and derive the companion masses. However, in some special cases the optical echelle
spectroscopy would not work. OGLE-TR-82 is one example, and becomes an interesting
case study.

After a thorough up study, in this paper we discard the planetary
nature of OGLE-TR-82, concluding that it is a blend with a reddened distant giant.
Section 2 describes the observations and reductions for all the data:
optical and infrared photometry and infrared spectroscopy.
Section 3 presents the results and discussion. The conclusions are listed in
Section 4.

\section{The Data}

This paper uses data obtained at various facilities, as listed in Table 1, which we
discuss in turn.

\subsection{Infrared Observations: Photometry and Spectroscopy}

OGLE-TR-82 was observed during the nights of May 4 and 5, 2006, using
SofI (Son of ISAAC; Moorwood \etal 1998) infrared camera and spectrograph
at the ESO NTT. SofI is equipped with a Hawaii HgCdTe detector
of 1024$\times$1024 pixels, characterized by a $5.4$ $e/ADU$ gain, a readout noise 
of 2.1 ADU and a dark current of less than $0.1$ $e/sec$. We used it
in  the large field camera mode, obtaining $4.9\times 4.9$ $arcmin^2$ field.
All measurements were made through the $K_s$-band filter (center = 2.162 $\mu$m
and width = 0.275 $\mu$m).

The reductions were made using IRAF tasks\footnote{IRAF is distributed by the National Optical Astronomy Observatories, operated by Universities for Research in Astronomy, Inc.}. 
First of all, the crosstalk 
correction was applied, taking into account the detector sensitivity difference
between the upper and lower half. Then the sky subtraction was applied by subtracting an 
appropriately scaled mean sky, which was obtained by combining 
all the offset images (two per target using ``dither-5'' mode) but 
rejecting outlying pixels.
Finally, we applied flat-field corrections to all images and aligned them. For the flat-fields, we
used the correction images provided by the NTT SciOps team\footnote{www.ls.eso.org/lasilla/sciops/ntt/sofi/reduction/flat\_fielding.html},
and the alignment was done with $lintran$ and $imshift$.

OGLE-TR-82 is listed as 2MASS source 10580297-6134263, with 
$J=13.479\pm  0.059$,
$H=12.077\pm  0.035$, and
$Ks=11.538\pm 0.030$.
The 2MASS yields a very red color for this star: $J-K=1.94$.

OGLE-TR-82 is located in Carina, in the Galactic plane at $l, ~b= (289.8638^o, ~-1.6131^o)$.
This region is relatively crowded, hence we find that for some OGLE candidates
the 2MASS photometry differs from ours.
The 2MASS photometry may be contaminated by the nearby neighbours
due to the pixel scale of 2MASS. We obtained the $K$-band magnitude by
using deeper and better sampled SofI images. 
The calibration of our infrared photometry was made using 2MASS.
The zero point of the $K$-band photometry should be good to 0.1 mag, which
is accurate enough for our purposes.

Using aperture photometry with an aperture small enough to
exclude nearby neighbours, we find $K=12.20\pm 0.10$ for this star.
Even though this is more than half a magnitude fainter than the 2MASS value,
in what follows we will adopt this as the final $K$ magnitude for OGLE-TR-82.
This allows us to plot OGLE-TR-82 in Figure \ref{uno}, which shows the 
periods $vs$ the $I-K$ color for several OGLE transits 
candidates observed in the Carina fields.  OGLE-TR-82 was clearly unique in 
this sample, as a very red object with a short period companion, that warranted
further study.

Therefore, we also acquired near-IR spectrum with SofI, covering the
region from 1 to 2.5 microns. The low-resolution blue and red grisms
delivered spectral resolution R$\sim$700, with slit 1 arcsec. This
is adequate for spectral classification. The total integration was
480 sec in the blue and 720 sec in the red, split into individual
integrations of 240 sec and 180 sec, respectively. The usual
observing technique of nodding along two slit positions was used.
The data reduction included the following steps: cross talk removal,
flat fielding, subtraction of the sky emission, extraction of
1-dimensional spectrum from each individual image, wavelength
calibration and combination into a final spectrum for each mode
separately. Finally, we removed the telluric absorption observing a
solar analog Hip 54715 (see for details Maiolino \etal 1996).
The two SofI spectra were joined together by scaling of the overlapping
regions from 1.51 to 1.645 microns and the spectra were not flux calibrated.

\subsection{VLT Optical Observations and Photometry}

The $V$-band VLT observations and photometry are described by Fern\'andez et
al. (2006).  Briefly, the photometric observations were taken with
VIMOS at the Unit Telescope 4 (UT4) of the European Southern
Observatory Very Large Telescope (ESO VLT) at Paranal Observatory
during the nights of April 9 to 12, 2005.  The VIMOS field of view
consists of four CCDs, each covering 7$\times$8 arcmin, with a
separation gap of 2 arcmin, and a pixel scale of $0.205 ~arcsec/pixel$.
We used the Bessell $V$ filter of VIMOS, with $\lambda_0=5460{\rm \AA}$, $FWHM=890{\rm \AA}$. 

A number of OGLE transit candidates were monitored simultaneously.
OGLE-TR-82 was located in one of the four monitored fields, and it
was scheduled to have a transit during the first night of our
run. However, a later revised ephemerides provided by OGLE
(A. Udalski private communication) showed that the transit ocurred
after the night was over. Nonetheless, some of the VIMOS images
had exquisite seeing, and we can use them for precise astrometry
and to measure the magnitude
of OGLE-TR-82 in the presence of a brighter optical neighbour located $1$ arcsec away.
It was necessary to do PSF photometry using DAOPHOT in IRAF in
order to account for this brighter nearby neighbour.

The best seeing images ($FWHM=0.5~ \mathrm{arcsec}$) taken near the zenith were
selected, and a master optical image was made.
The images of OGLE-TR-82 analyzed here are 400$\times$400 pix,
or 80 arcsec on a side.  This small image contains about 500
stars with $15<V<24$.
Figure \ref{4juntas}-(a) shows a $20\times 20$ arcsec$^2$ 
portion of this  $V$-band master image, showing OGLE-TR-82 with a
larger circle, and its neighbour located 1" North with a smaller circle. We measure
$V=20.61\pm 0.03$ for OGLE-TR-82, and $V=17.388\pm 0.01$ for its neighbour. 
The OGLE $I$-band image is shown in Figure \ref{4juntas}-(b) for comparison. 
both stars have similar magnitudes in this image. Taking $I=16.30\pm 0.10$
from Udalski \etal (2002), we obtain a very red color $V-I=4.1\pm 0.1$
for OGLE-TR-82. This color indicates that the target is a very late
M-dwarf or an early brown dwarf. 
The $Ks$-band image of Figure \ref{K} shows that the relative brightness of both stars
is reversed: OGLE-TR-82 is much brighter than its neighbour in the near-IR.

\subsection{GEMINI-South Optical Observations and Photometry}

The object is so faint in the optical passbands, that it takes an 8-m telescope 
to  measure the transit accurately. Therefore,
a full transit of  this target was observed with the GMOS at the GEMINI-S
telescope. The observations were acquired in queue mode on 27 January 2006.
The GMOS field of view is $5.5'\times 5.5'$, with a scale of $0.0727$ arcsec/pix.
This scale is finer  than that of VIMOS, but the seeing during the observations 
was worse than the VIMOS run, which complicates the transit
photometry in the presence of the brighter neighbour $1$ arcsec away. 
The beginning of the transit was noisier because of weather problems,
but the night stabilized after the ingress, resulting in a relatively good light curve.

\begin{figure}
\begin{center}                                            
\leavevmode
\plottwo{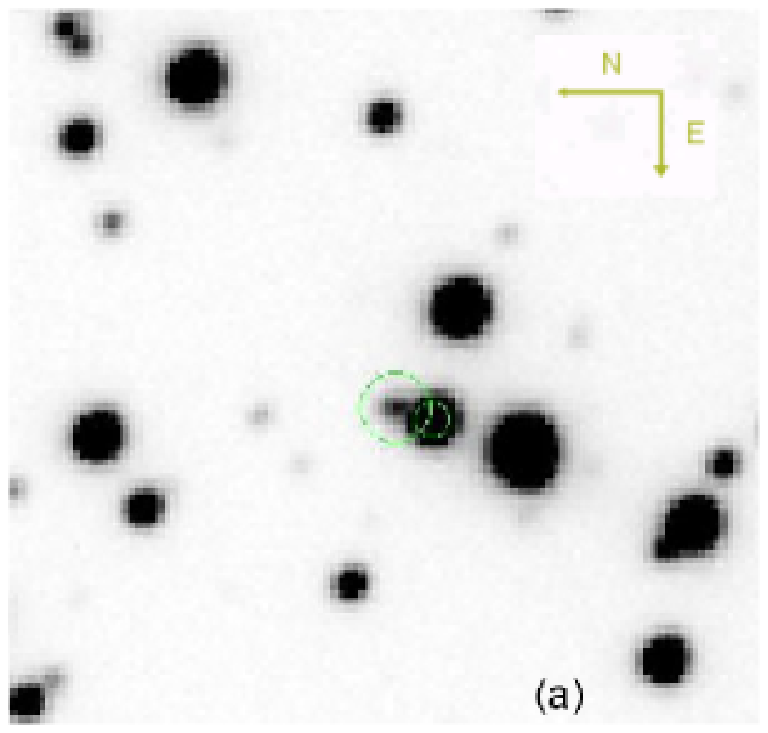}{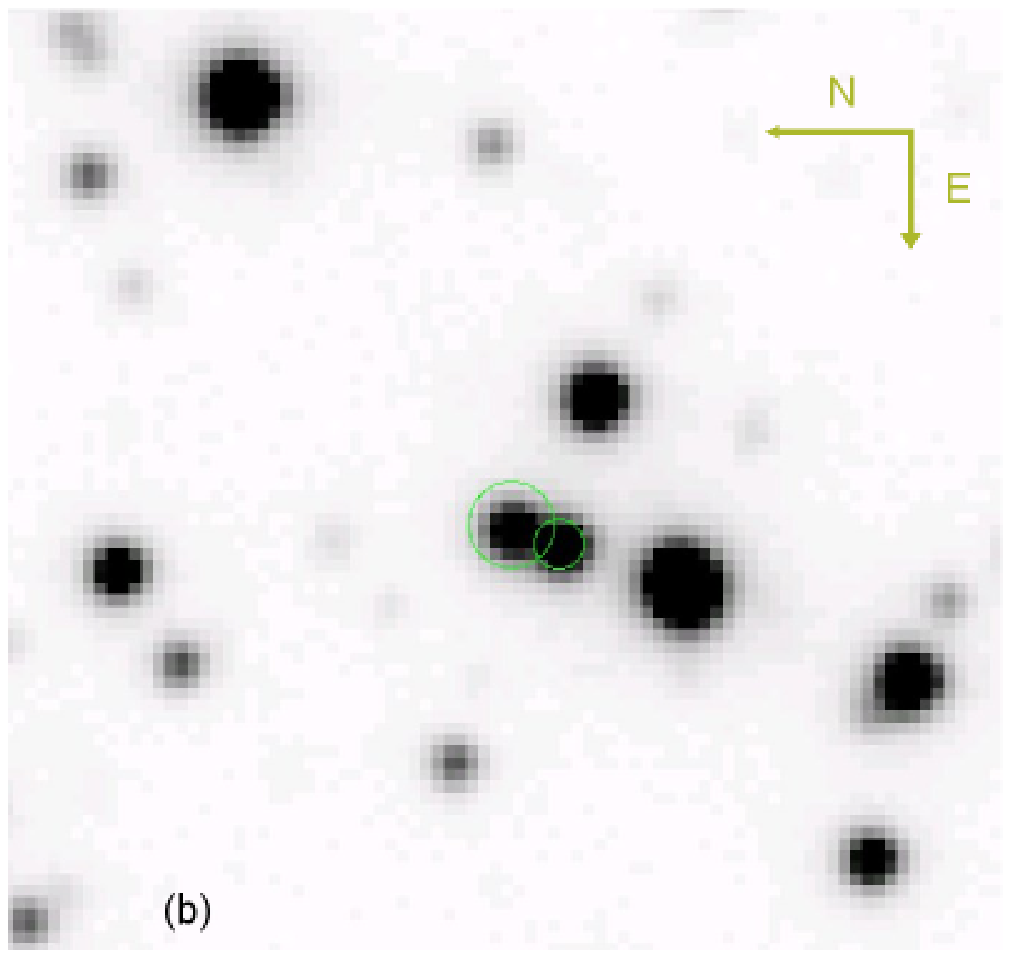}           
\plottwo{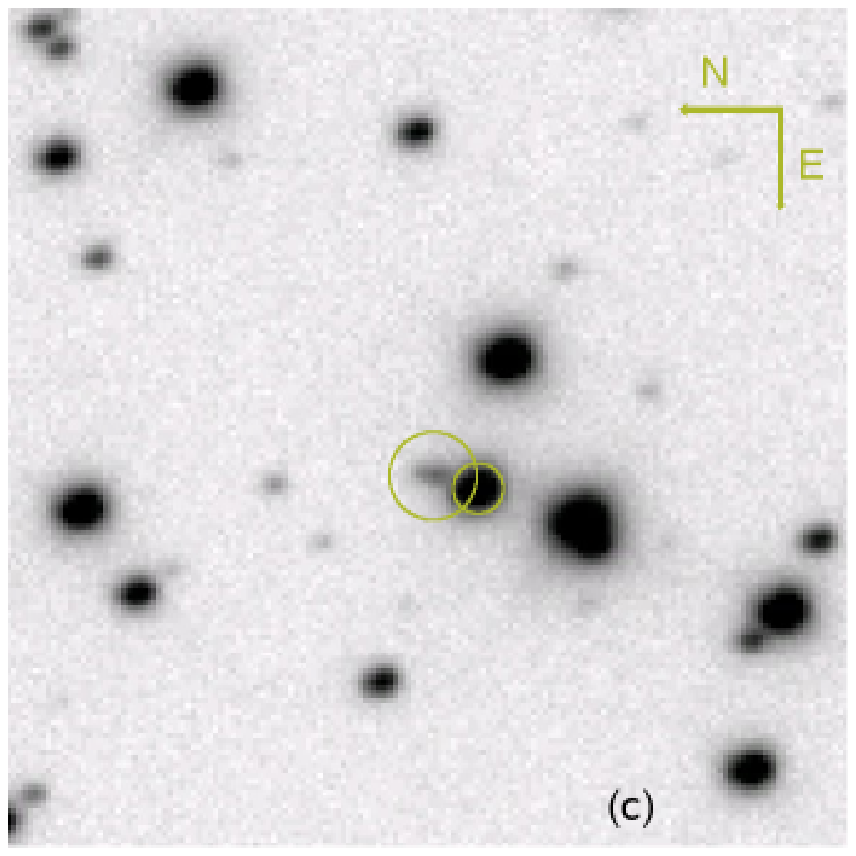}{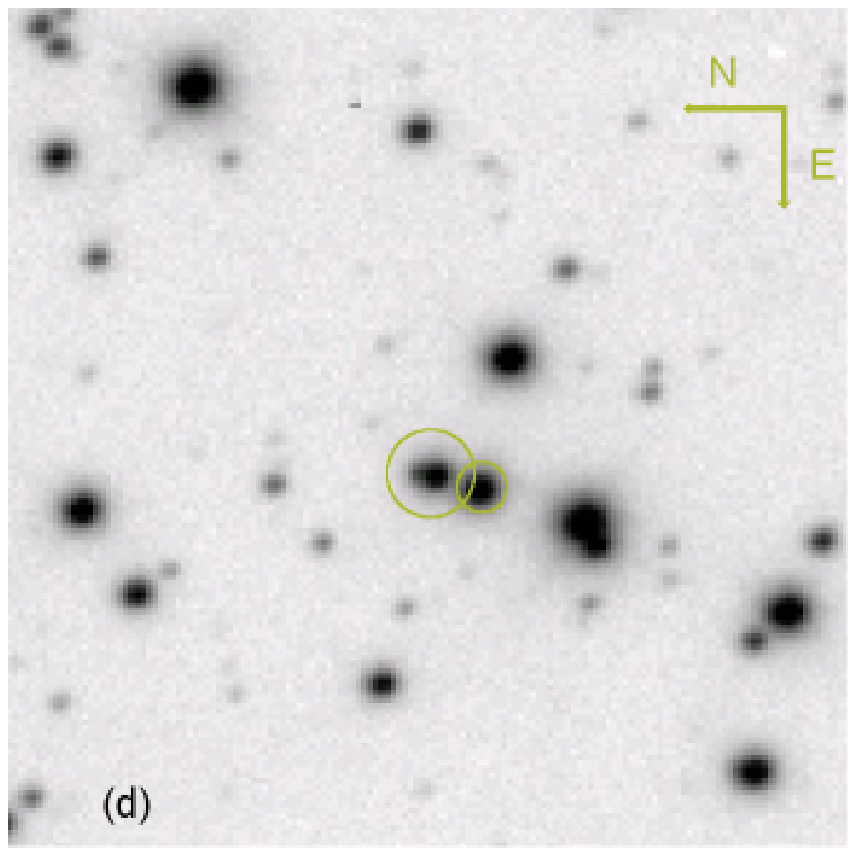}       
\end{center}
\caption{ 
Portions of images in differents bands including OGLE-TR-82 which is the star at the center of the big circle.
These images covers $20\times 20$ arcsec.
(a): 0.5 arcsec seeing VIMOS $V$-band image.  OGLE-TR-82 has $V=20.6$ and the faintest stars seen have $V\sim 24$. The size of the large circle is approximately the size of the fibers
used in previous attempts that failed to measure radial velocities for this candidate.
Under normal $1$ arcsec seeing conditions, this target is swamped by the
nearby neighbour. 
(b): 0.85 arcsec seeing OGLE $I$-band image.  OGLE-TR-82 has $I=16.30$ and the faintest stars seen have $I\sim 22$. 
(c): 0.5 arcsec seeing GEMINI-South $g$-band image. The faintest stars seen have $g\sim 24$.
(d): 0.5 arcsec seeing GEMINI-South $i$-band image. Compare this high resolution image with the OGLE-finding chart shown in (b).
}
\label{4juntas}
\end{figure}

\begin{figure}
\plotone{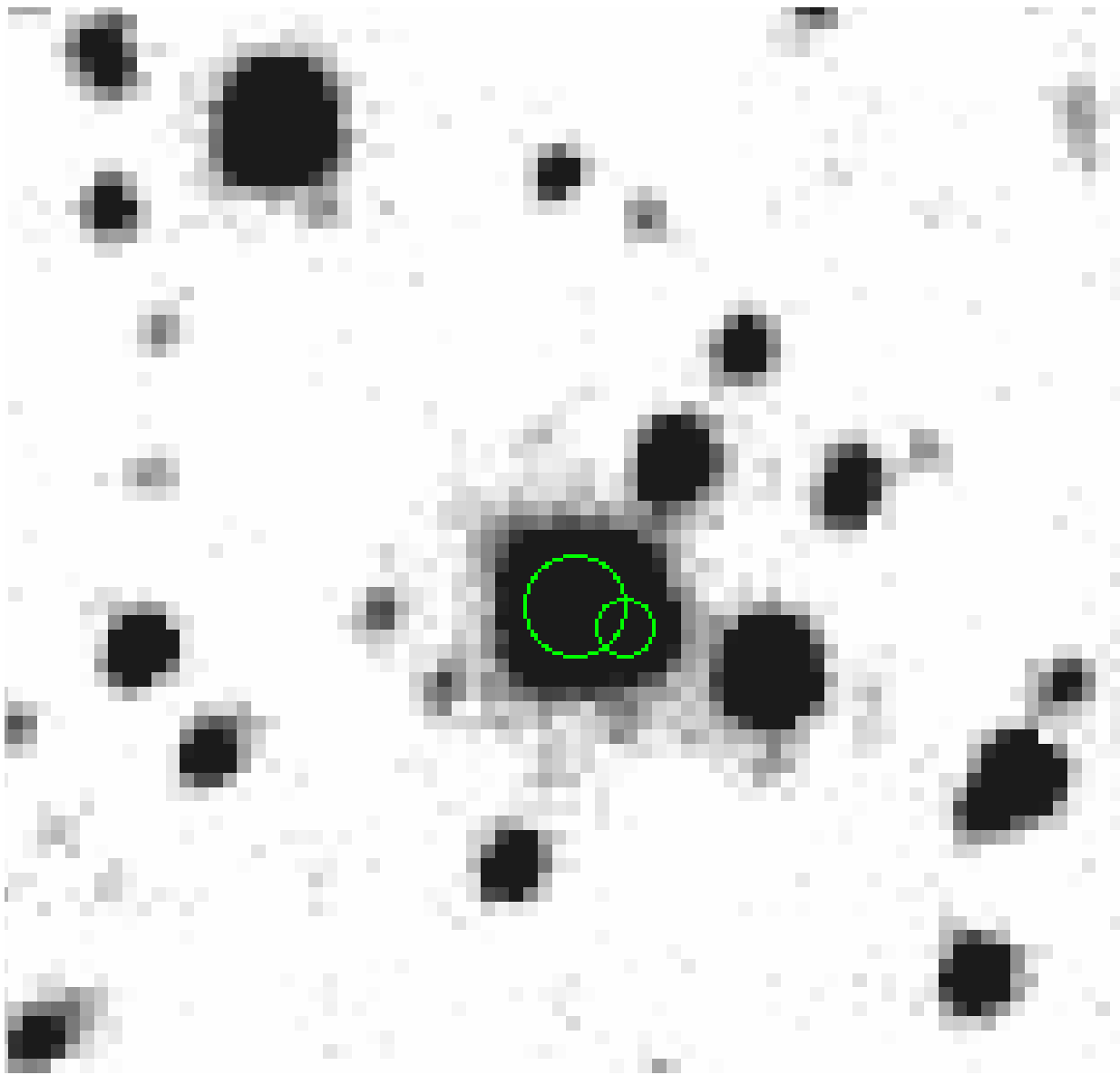}
\caption{
Portion of a 0.7 arcsec seeing SofI $K$-band
image including OGLE-TR-82 ($K=12.2$),
which is the bright star at the center of the large circle. This image covers
$20\times 20$ arcsec as in Figure \ref{4juntas}  (scale $0.287 ~arcsec/pix$), 
and the faintest stars seen have $Ks\sim 18$.
}\label{K}
\end{figure}

We used the $g$ and $i$-band filters (GEMINI filters g-G0325 and i-G0327, 
respectively), alternating 3 consecutive exposures of each filter in turn. 
This observation sequence was repeated without interruptions for a period
of 5.3 hours. There were 117 images for each of the two filters.
The seeing and image quality were not optimal, but they
were good enough for the difference imaging photometry.
Figures \ref{4juntas}-(c) and \ref{4juntas}-(d) show the best GEMINI-South 
single images of the OGLE-TR-82 field for comparison with the previous
Figures \ref{4juntas}-(a) and \ref{4juntas}-(b). The stellar images are elongated, but obviously of higher
resolution than the VLT and OGLE images. It can be appreciated that the source
has a faint unresolved companion 1-2 pixels away to the East.

\begin{figure}
\plotone{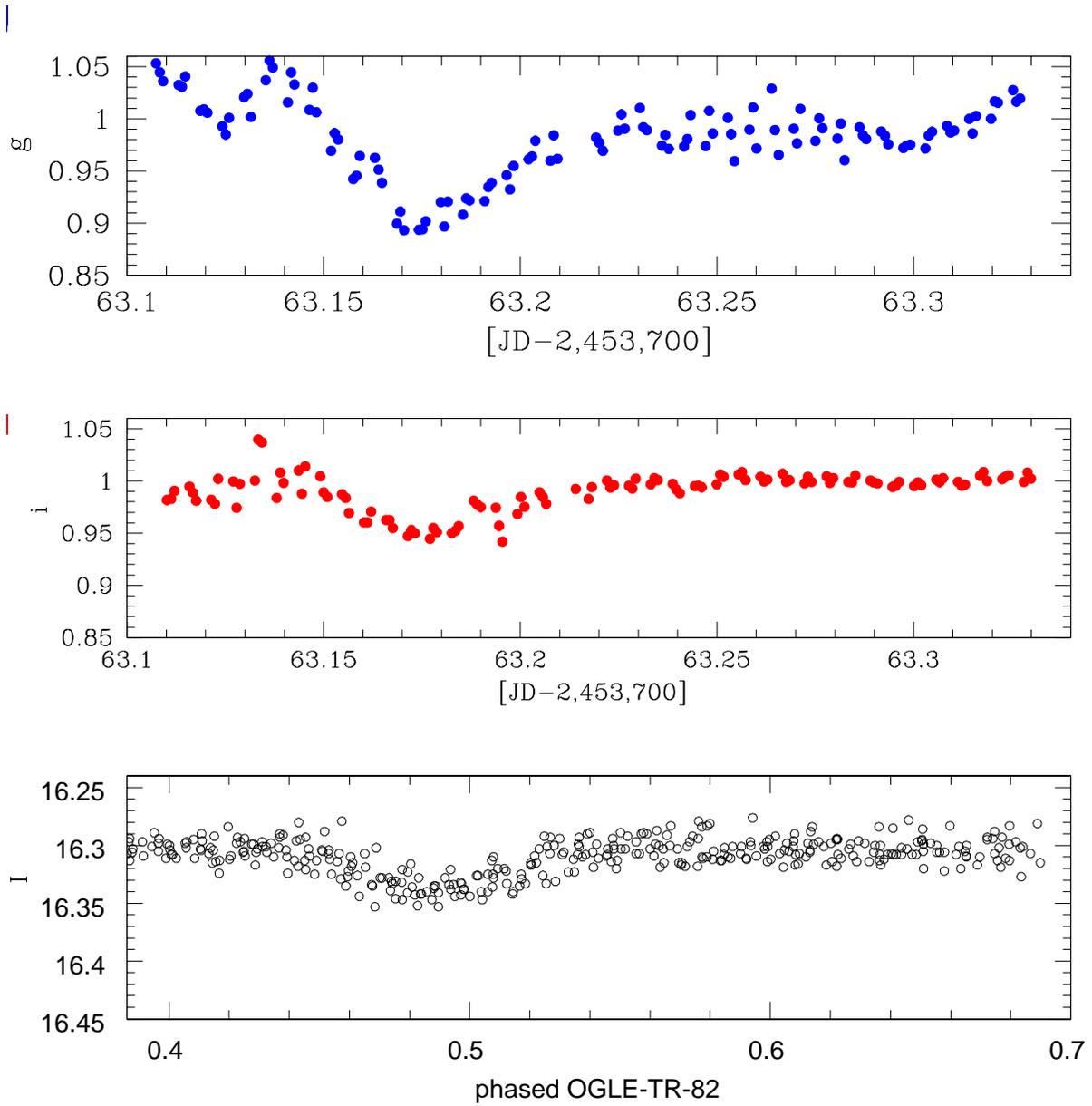}
\caption{
Single transit of OGLE-TR-82 observed with GMOS at the GEMINI-South telescope in the
$g$-band (top) and in the $i$-band (middle), compared with the OGLE $I$-band data phased 
using $P=0.76416 ~d$ shown to the same scale (bottom).  
}\label{lc}
\end{figure}

The transit light curves were measured following the procedure described for
the VLT transits (see Fern\'andez \etal 2006).
The $g$-band transit was very difficult to measure, because the star
becomes too faint, and it is overwhelmed by the nearby neighbour.
There is a factor of 10 difference in counts in each individual image 
for this star between the $g$ and $i$-bands.
Figure \ref{lc} shows the full $g$ and $i$-band light curve for the GEMINI-South
observations, when the OGLE-TR-82 transit was monitored.  
There are about $30$ points in each of the light curves
during in our single transit shown in Figure \ref{lc}, 
and the minimum is well sampled, allowing us to measure accurate amplitudes.  
The resulting light curves (which were not corrected for a linear trend), yield
$rms_g=0.005$ and
$rms_i=0.002$ mag in the flat portion at the end of the night.
Figure \ref{lc} also shows the phased light curve of the OGLE
$I$-band photometry (in a similar scale) for comparison. The transit
is well sampled in our $i$-band observations, and the scatter appears smaller.  
We confirm                 
the amplitude of the transit measured by OGLE in the $I$-band, $A_i=0.034 ~mag$,
but find that the amplitude in the $g$-band is larger, $A_g=0.10 ~mag$.
Already the GEMINI-S data cast doubts on the planetary transit scenario, because
of the amplitude difference between the $g$ and $i$-band transits.
Also, the $g$-band light curve shows a more triangular eclipse shape, which
is more characteristic of a grazing binary.
However, the $g$-band light curve was hard to measure, and we cannot
discard residual contamination from the neighbours or fluxing problems
with the difference image photometry.

\subsection{Previous Spectroscopy}

\hspace{20pt} As noted in the Introduction, Pont \etal (2004) observed OGLE-TR-82 with 
UVES/FLAMES at the VLT, but they could find no CCF signal, concluding that this
was an unsolved case.  They point out that the object could be too red,
and suggested further observations. They suggest a K7/M0 dwarf, less
extreme than the spectral type found in the next Section.

We conclude that there was no signal for this object in the spectroscopic
observations because they were made in the $V$-band portion of the spectrum, 
where the target is simply too faint with $V=20.6$. Using the setup of Pont
\etal (2004), this target cannot give useful signal to measure velocities with
an accuracy of $\sim 30 ~m/s$ in any reasonable amount
of integration time according to the 
UVES/FLAMES exposure time calculator.
Having answered the mystery about the absence of spectroscopic signal 
for OGLE-TR-82, we open a more important question: is this star a giant or a dwarf?

\section{A planetary transit?}

With the present optical and infrared photometry one can estimate
some of the stellar parameters: the spectral type, luminosity, mass,
radius and distance.

\subsection{The Period of OGLE-TR-82-b:}

The VLT observations did not show a transit, which was scheduled for the end
of the night according to the original OGLE ephemeris. 
Udalski \etal (2005, private communication) revise the ephemeris for
this object, giving mean transit times of:
$$JD=2452323.84747 + 0.764244  t.$$
This puts the transit beyond the end of the 
VIMOS observing night, explaining its absence.
At the same time, this is a minor revision that
allowed us to recover the transit for the GEMINI observations.
With the transit observed at GEMINI-S, the ephemeris is improved. We
obtain a period similar to OGLE: $P=0.7643813\pm 0.0000010$ days.

\subsection{The Spectral Type of OGLE-TR-82-b:}

In a previous work, we have used optical and infrared photometry to
characterize OGLE extrasolar planetary companions (Gallardo \etal 2005).
We can estimate the spectral type of OGLE-TR-82 first using the multicolor photometry.
The stellar parameters from OGLE and the present photometry are listed in Table 2.

\begin{figure}
\plotone{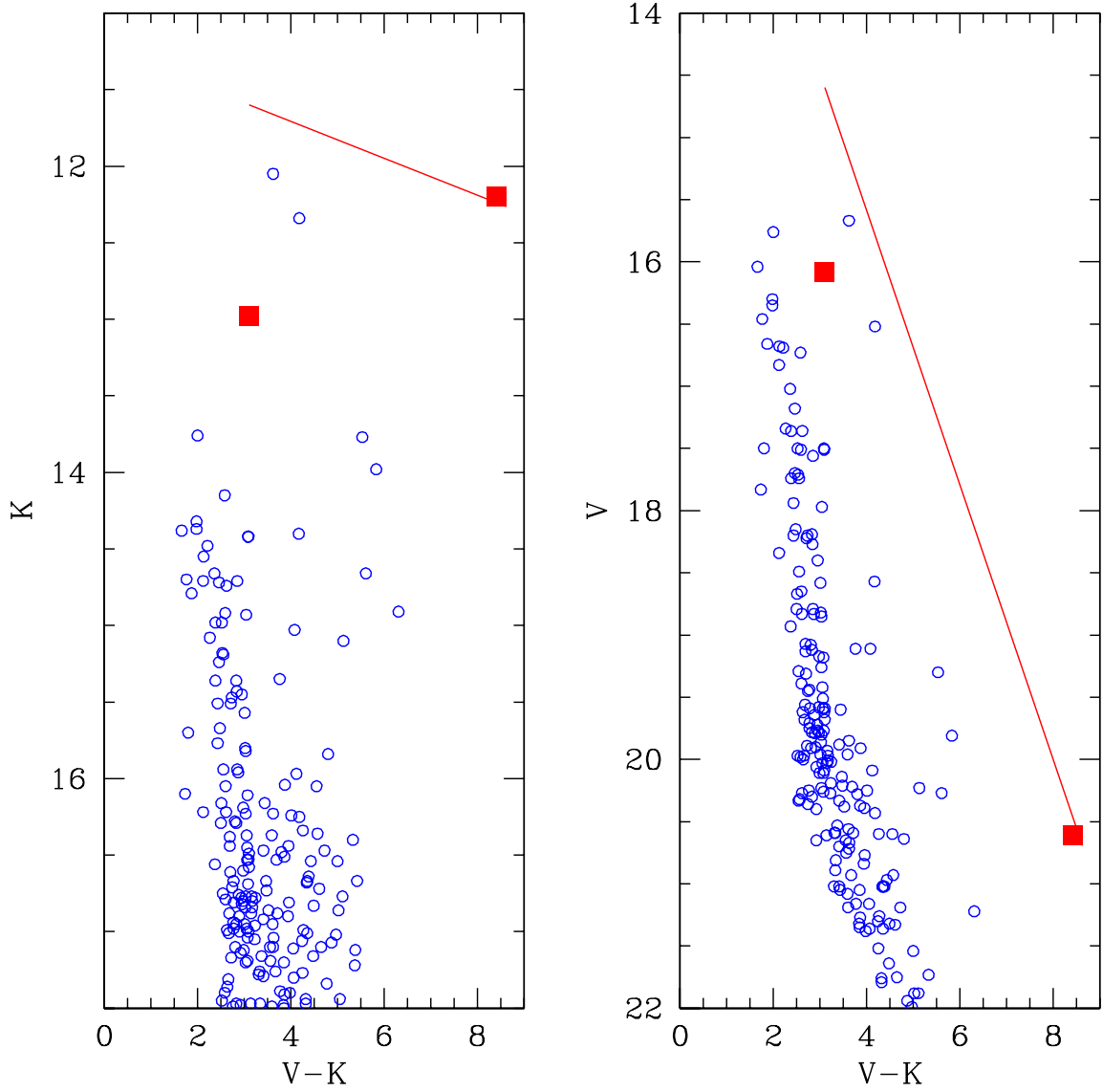}
\caption{
Optical-infrared color-magnitude diagrams of all stellar objects in a $1'\times 1'$ field in Carina.
In both panels, the big square on the right is OGLE-TR-82, while the one on 
the left is OGLE-TR-113 shown for comparison.
The reddening vector corresponding to $A_V=6 ~mag$ is shown for OGLE-TR-82.
}\label{cm}
\end{figure}

The optical-infrared color-magnitude diagrams are shown in Figure \ref{cm}
for all stars in a Carina field of
about $1'\times 1'$. The disk main sequence is
very well defined. The target star OGLE-TR-82 is located away from this main sequence,
indicating that either this is a nearby late $M$-dwarf, or a very reddened distant giant. 
Because both alternatives are very
different from the spectral type of K7-M0V adopted by Pont \etal (2004),
we consider these two possibilities that arise from the photometry.

\begin{figure}
\plotone{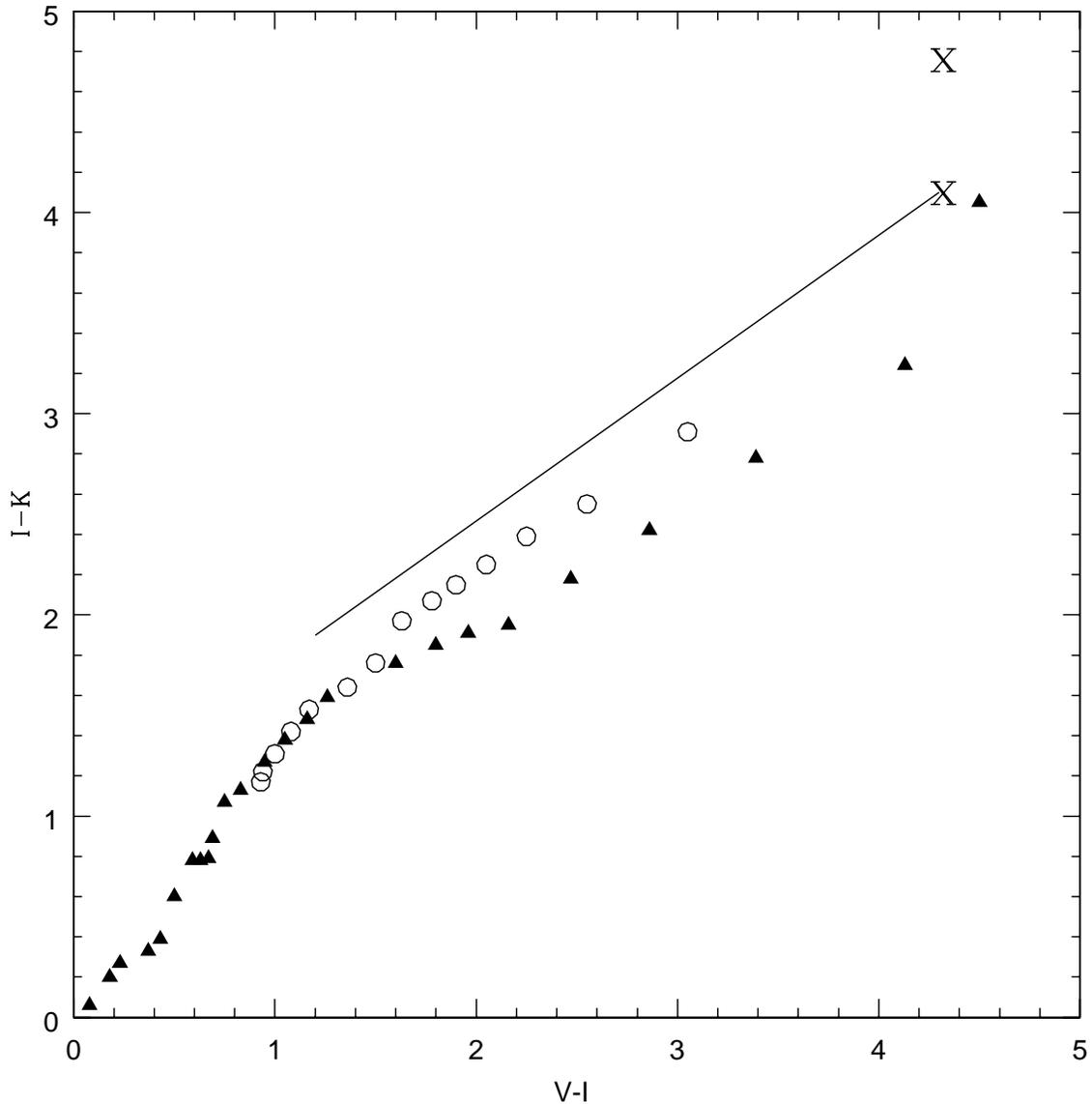}
\caption{
Color-color diagram indicating the fiducial loci of giants (open circles)
and dwarfs of different spectral types (full triangles).  The observed position 
of OGLE-TR-82 is marked with the crosses. The top one comes from the
2MASS colors, and the bottom one comes from our photometry. This position
is consistent with a late main-sequence star, next to the location of 
a typical M7V star. The direction of the reddening vector is indicated
with the straight line, the length gives the vector corresponding to a total
$A_V=6.0 ~mag$.  Considering this reddening, the star would lie closer to the 
location of K-type red giants.}\label{cc}
\end{figure}

Figure \ref{cc} 
shows the loci of giants and dwarfs of different spectral types
in the $V-I$ $vs$ $I-K$ color-color
diagram. The position of OGLE-TR-82 in this diagram is consistent with an
unreddened spectral type of M7V,
with an error of about one subtype (between M6V and M8V).
For this spectral type the mass and radius are 
$M_s=0.10\pm 0.02 ~M_{\odot}$, and $R_s=0.15\pm 0.02 ~R_{\odot}$.
This star lies at the
boundary between M-dwarfs and brown dwarfs. In fact, adopting the 2MASS photometry
($V-K=9.1$) instead of our values ($V-K=8.4$) results in an L-type brown dwarf.
The spectral type of M7V would yield an absolute magnitude of $M_V=18.60$
(Bessell 1991). The resulting distance modulus of $m-M=2.01$ gives a
distance of 25 pc, and would make this the nearest OGLE transiting planetary candidate.
Most other OGLE transiting candidates are located at distances between a few
hundred parsecs and a few kiloparsecs (Gallardo \etal 2005).

However, Figure \ref{cc} also shows that if the reddening is severe, $E(B-V)\approx 2 ~mag$, the
location of OGLE-TR-82 can also be consistent with a distant red giant.
\subsection{The Radius of OGLE-TR-82-b:}

If we consider the case of an M7V dwarf, the OGLE $I$-band 
transit light curve amplitude yields $R_p/R_s = 0.18$.
This is a small planetary radius, similar to the radius of Neptune,
$R_p=0.3 ~R_J$.

This is very different from the two previously published values 
for the radius of the OGLE-TR-82 companion,
both based on the OGLE photometry. Udalski \etal (2002) obtain
$R_p=1.1 ~R_J$, and
Silva \& Cruz (2006) assume a star with $M= 0.65 ~M_{\odot}$, and obtain
a radius $R_p= 1.10 ~R_J$.
Both values are much larger than the radius
implied for a M7V star, indicating the dramatic change due to the 
new spectral type obtained for OGLE-TR-82 based on the $V-K$ color of this source
if it is indeed a main sequence star. 

The present photometry shows that the previous 
attempts to measure the radial velocities of OGLE-TR-82 failed because
this object is much too faint in the optical. It is now clear
that with $V=20.61$ it would have been impossible to measure. 
However, velocities in
the near-infrared are within reach of the largest telescopes, because
the target is very red $K=12.2$. If this were a low mass
star with a planet in a tight orbit, these velocity measurements do not need 
to approach the few m/s accuracy as usually needed for the RV planet searches.
In this case, a Jupiter-mass planet would yield a 
radial velocity semiamplitude of about $1.1 ~km/s$,
well within reach of near-infrared spectrographs at large 8m class telescopes.
A Saturn-mass planet would give a smaller radial velocity semiamplitude of $300 ~m/s$.

The smallest extrasolar planet currently known is
HD149206-b, with $R_p=0.7 ~R_J$ (Sato \etal 2005),
a Saturn-mass planet with a large dense core.
The smallest companions so far have been detected in the bulge
fields by Sahu \etal (2006) around stars of $0.5 ~M_{\odot}$.
If OGLE-TR-82 were an K7-M0V star (Pont \etal 2004), it
would be an extreme case because of its low mass, small size, 
and short period companion. The planetary option would be still open for this target
if we could confirm the M-dwarf of later type.

In the absence of spectroscopic velocities, we cannot estimate the
mass of the companion. However, in order to give an idea of the
possibilities, we note that a companion with $M=1 ~M_J$ would give
an orbital semimajor axis of $a = 10.8$ and $R_s = 0.0076 ~AU$. 

There is, however, a problem with this interpretation: the total transit time
observed is $t_T\approx 0.04 ~d$, while the stellar and orbital parameters would predict
a shorter transit of $t_T=0.027 ~d$. 

Though interesting as a good planetary candidate at this stage, a deeper study
shows that the interpretation is not so simple, as discussed below. 

\subsection{A triple system?}
We have explored different possibilities for the nature of 
the OGLE-TR-82 companion: 
a planetary companion to an M-dwarf,
a white dwarf transiting in front of a M-dwarf,
a brown dwarf transiting in front of a M-dwarf,
a grazing M-dwarf tight binary of similar masses, variability like spots,
a main sequence star transiting in front of a red giant, 
and a blend with a reddened background giant star.
The last possibility is the only one that agrees with all the available data, 
as discussed below.

\subsection{Blend with a background star hypothesis:}

An eclipsing binary blended with a background star yields low amplitude transits
because the light of the contaminating star dilutes the eclipse.
Alternatively, the eclipses of a distant binary would be diluted by a blend
with a foreground star.

It is now well known that transit searches are plagued by blends, and
that in order to confirm the planetary nature of a transit candidate 
the radial velocity orbit must be obtained. In the present case the
blend option appears the most likely.

The best seeing optical images of GEMINI-South (Figures \ref{4juntas}-(c) and \ref{4juntas}-(d) shows an elongated shape even for the faint OGLE-TR-82 star. This is obvious in the $g$-band (Figure \ref{4juntas}-(c)), but less evident in the $i$-band (Figure \ref{4juntas}-(d)). 
The elongation is in the E-W direction, but the pair is faint and unresolved, we estimate the separation to be approximately 1-2 pixels in the GEMINI-South images.
One of the sources appears to be the very red source (to the N), while the
other (to the S) shows normal color of typical main sequence field stars.
Note that spectroscopic fiber sizes (e.g. HARPS, FLAMES) would not be able
to resolve the light form both stars, the spectra would necessarily be the
composite spectrum of the blend.

We conclude that
the red source is a reddened background giant, while the other source is a
normal main-sequence binary. The blend reduces the amplitudes of the
eclipses of the binary as seen in the $I$-band, mimicking the light curve of
a planetary transit.

It is usually not possible to distinguish between the cases of a
foreground binary blended with a background star $vs.$ a background
binary blended with a foreground star.
In the present case, while a 
late main-sequence star transiting in front of a red giant
can produce an eclipse of the measured depth $A=0.034 ~mag$,
a blend of a foreground main sequence star with a background binary
red giant does not work, because the measured transit time of 1.5 hr 
is too short compared with typical red-giant sizes, even taking into account
the short period of OGLE-TR-82. It would
have to be a grazing eclipse with a marked triangular shape.

The final solution to the dilemma comes from near-IR spectroscopy obtained
with SofI at the NTT. 

\begin{figure}
\plotone{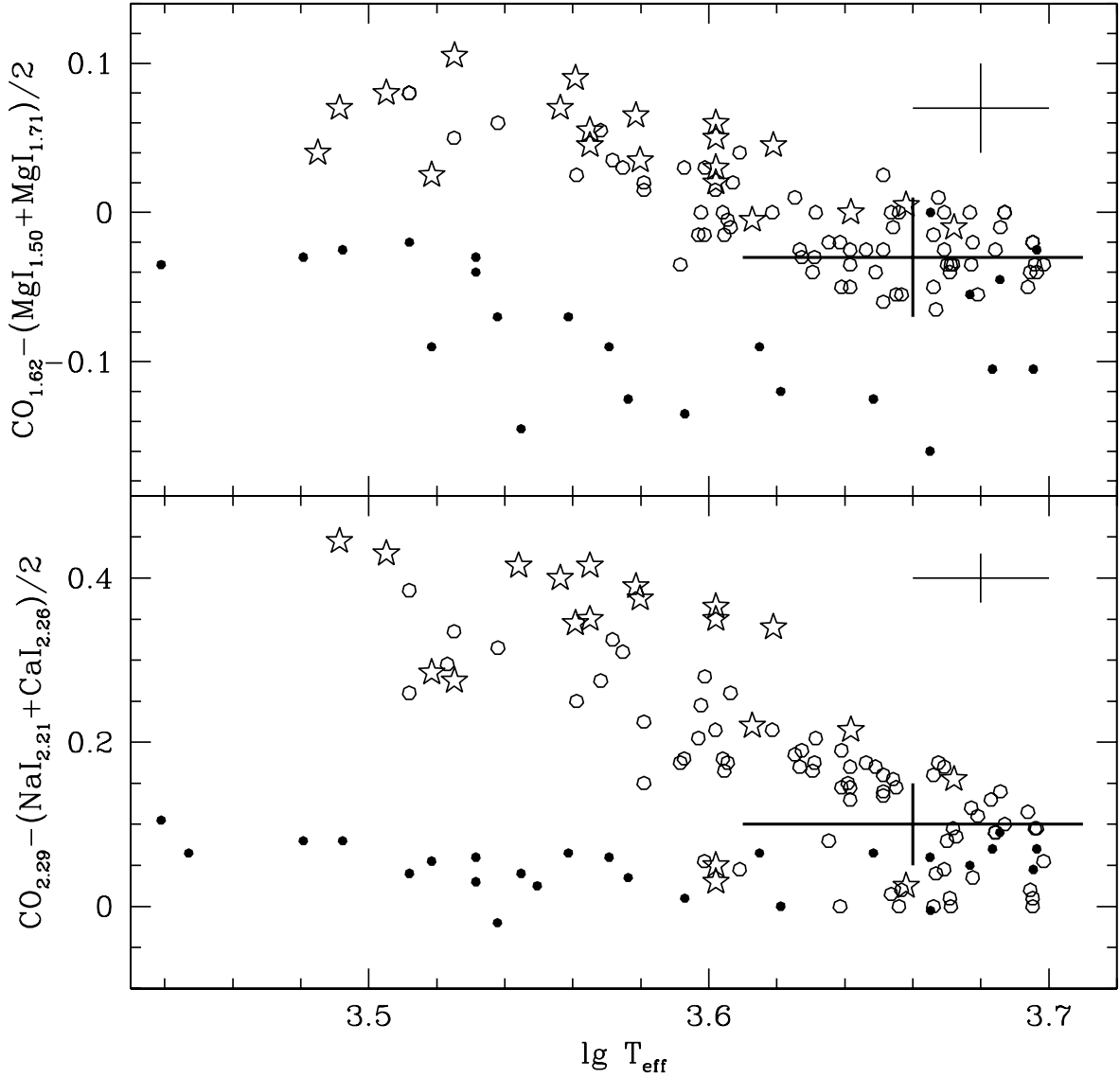}
\caption{
Two-dimensional spectral classification with H-band features (top) and K-band features (bottom). All indices are in magnitudes.  Star symbols indicate supergiants, open circles are giants, and solid dots represent dwarfs and subgiants. The typical ±1 measurement error is shown in the top right corner. The CO bands are defined by Origlia \etal (1993), the Ca and Na indices are from Ali \etal (1995), and the Mg definitions are from Ivanov \etal (2004).
}\label{lumclass}
\end{figure}

The first question that our near-IR spectra allow us to address is
if the star is a giant or a dwarf. The 2-dimensional spectral
classification is discussed in detail in Ivanov \etal (2004, see
Section 5.2). First, we used Eqs. 2 and 4 to determine the T$_{eff}$
and then placed the object on the plots shown in Figure 12 (left) in
Ivanov \etal (2004) (Figure \ref{lumclass}). We measured Mg{\sc i} 1.50\,$\mu$m, Si{\sc i}
1.58\,$\mu$m, CO 1.62\,$\mu$m and Mg{\sc i} 1.71\,$\mu$m indices as
defined in that paper on the extinction corrected spectrum to obtain
0.11$\pm$0.04 mag, 0.03$\pm$0.01 mag, 0.04$\pm$0.02 mag and 0.06$\pm$0.02 mag,
respectively. Note that here and further in this section the
uncertainties should be treated with caution because they include
only the Poisson errors from the spectra and not any systematic errors
that might rise from the sky subtraction or the telluric corrections,
for example. Also, we did not correct for any differences due to the
lower resolution of our spectra with respect with the library of
Ivanov et el. (2004). The derived stellar effective temperatures were
4077 and 5101\,K. We averaged them to obtain T$_{eff}\sim$ 4600 K,
suggesting the star has K0 spectral type. The {\it CO\,1.62 -
(Mg{\sc i}\,1.50 + Mg{\sc i}\,1.71)/2 vs log\,T$_{eff}$} plot (Fig.
\ref{lumclass}, top) firmly places the target
among the giants while the {\it CO\,2.29 - (Na{\sc i}\,2.20 +
Ca{\sc i}\,2.26)/2 vs log\,T$_{eff}$} plot (Fig. \ref{lumclass}, bottom) is inconclusive.

Finally, we applied the metallicity calibration technique of Frogel
\etal 2001 (see their Eq. 3) based on the strength
of Na{\sc i} 2.20\,$\mu$m, Ca{\sc i} 2.26\,$\mu$m and the 2.3\,$\mu$m
CO band. We preferred the spectroscopic method rather than the
combination of spectroscopic and photometric indices because our
target suffers strong extinction and even a small uncertainty in
the extinction would cause significant error in the metallicity
estimate. Following the prescriptions of Frogel \etal (2001) and
Ram\'{i}rez \etal (1997) we measured the following equivalent widths:
1.5$\pm$0.1, 1.6$\pm$0.1 and 10.2$\pm$2.0\AA, respectively for the
Na, Ca and CO features. For the metallicity estimates we tentatively
doubled the errors. This yields [Fe/H]$_{ZW}$=$-$1.0$\pm$0.1 in the
Zinn \& West (1984) metallicity scale. The transformation of Frogel
\etal (2001; Eq. 6) give [Fe/H]$_{CG}$=$-$0.86 in the Carretta \&
Gratton (1997) scale. Frogel \etal (2001) also give a metallicity
calibration based on the CO strength alone. It yields
[Fe/H]$_{ZW}$=$-$1.06$\pm$0.13, in agreement with the first estimate.

Of course, the classification criteria of Ivanov \etal (2004) and
the calibrations of Frogel \etal (2001) were derived for disks or
globular cluster stars with different age, chemical enrichment
history, abundance ratios, etc. than the Milky Way halo stars.
Therefore, our results should be treated with caution and we can
only conclude that the OGLE-TR-82 primary is consistent with being
a metal poor red giant.

In addition, we used the library of Pickles (1985) to carry out empirical spectral classification by comparing dwarfs and giants templates spectra with the near-IR spectrum of our target.  From this way, this spectrum first confirms the extremely red nature
of this object, as deduced from the IR photometry. Second, and most
importantly, the spectral type obtained is consistent with the K3III, with an estimated
absorption $A_V=7 ~mag$. 

\begin{figure}
\plotone{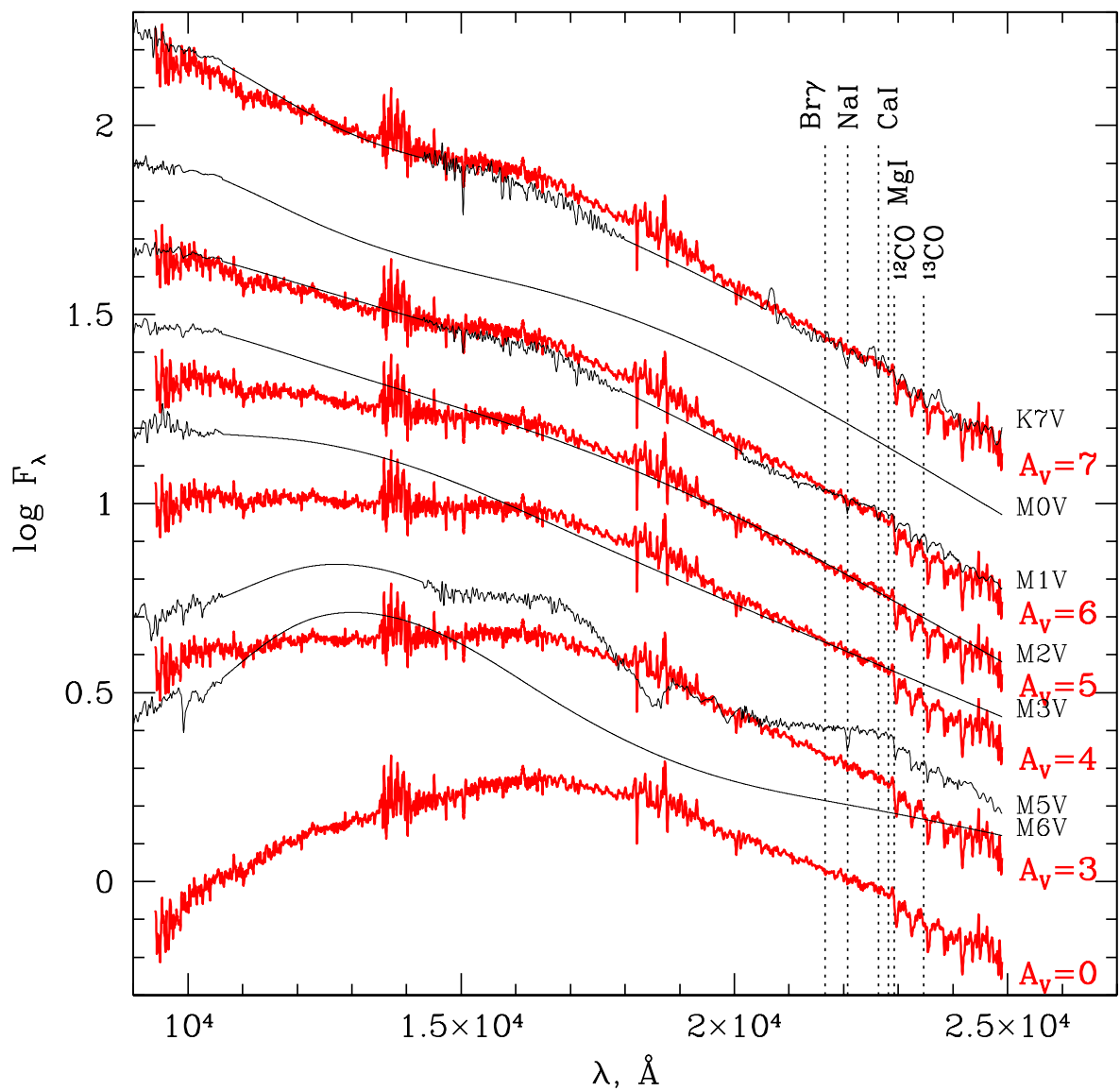}
\caption{
Near-IR spectrum of OGLE-TR-82 corrected for different reddening amounts (red thick curves)
compared with templates of dwarf stars from the spectral library of Pickles (1985) (black thin curves). The curves are shifted by a constant for a better comparison (y-axis is in arbitrary units).
}\label{esp1}
\end{figure}

The strong CO band at 2.3 micron makes it impossible to
match the host star with the spectrum of any red dwarf from the spectral
library of Pickles (1985), 
except for the latest spectral types M6-M7. However, the specific shape
of the continuum for these stars does not match the shape
of the continuum of our star (Figure \ref{esp1}). 
Earlier dwarfs than M5 are also inconsistent with the observed colors,
unless one assumes unrealistically larger absorption of
$A_V>12-15$ mag, which appears impossible for a nearby star.

\begin{figure}
\plotone{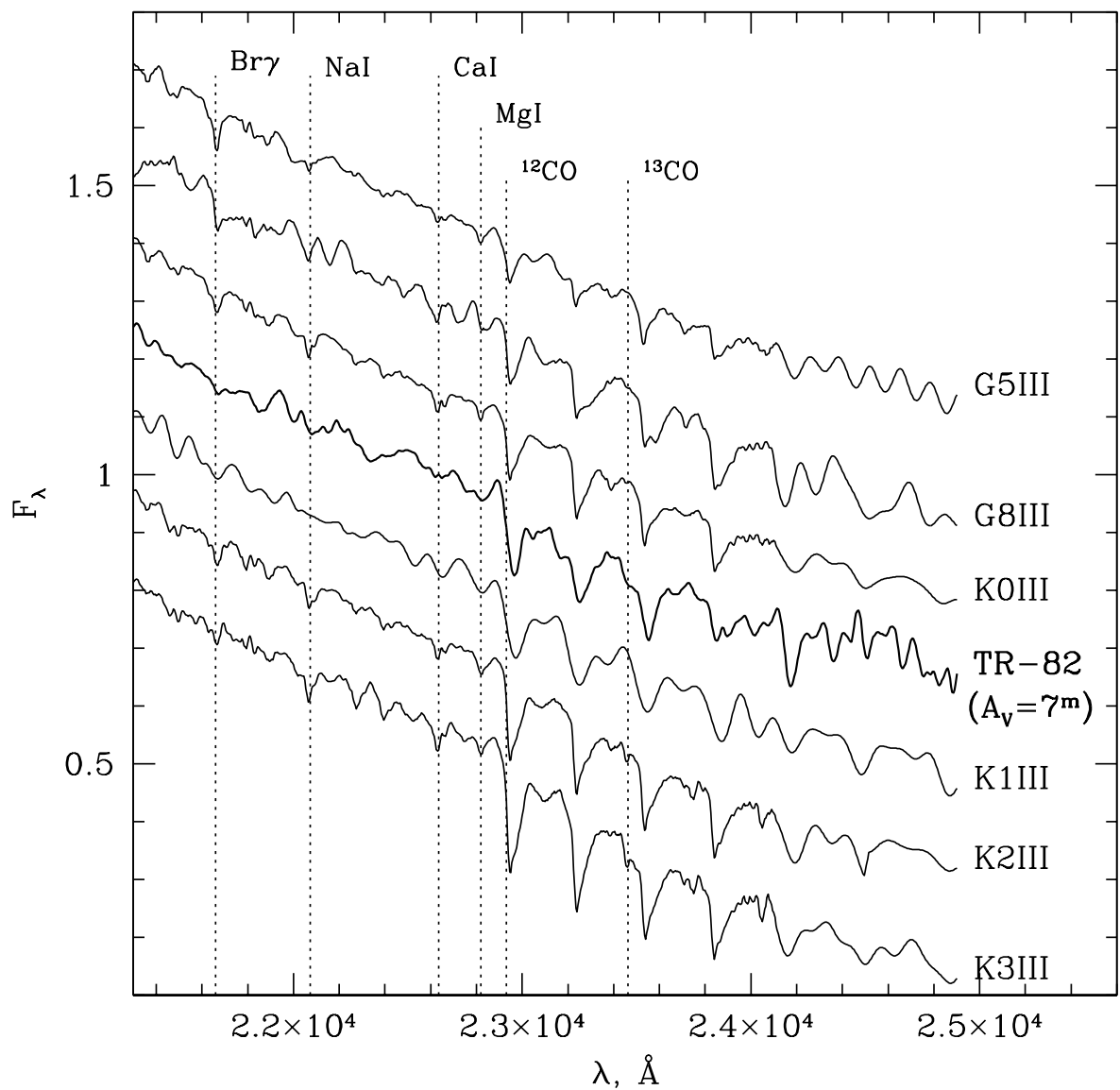}
\caption{ K-band spectrum of the OGLE-TR-82 primary (thick
line) compared with giant templates from the spectral library of Pickles (1998) (thin lines). 
Some more prominent features are marked. Given the signal-to-noise
and the resolution of the spectrum, only the stronger features are
detected. The curves are shifted by a constant for a better comparison (y-axis is in arbitrary units).
}\label{esp2b}
\end{figure}

This leaves us only with the possibility of a
distant giant star, supported by the shape of the continuum.
The comparison with the giant template spectra shown in Figure 
\ref{esp2b} 
shows that the best fits are found with
K3III giant and $A_V=7 ~mag$, as expected for a distant giant 
star located in the Galactic plane. 
Stars of later types present
poor matches because of the strong water vapor features.
We caution that some contamination from the nearby blue star is
expected in the blue part of the spectra at 1 micron.


This spectral type finally rules out a planetary size companion. Even though the mass
of the secondary is unconstrained due to the lack of radial velocities,
we conclude that this companion is most probably a late-type main sequence star. 
Thus, the OGLE-TR-82 system is an eclipsing binary,
blended with a background reddened K giant.

\section{Conclusions}

Udalski \etal (2002) discovered low amplitude transits in the main
sequence star OGLE-TR-82, which was considered as a prime planetary candidate
orbiting a M-type dwarf (Pont \etal 2004, Silva \& Cruz 2006). 
We find that this object has an extremely red color $V-K=8.4$,
making it unique among the OGLE transiting candidates in Carina. 
Future space based missions like COROT and KEPLER may discover low
amplitude transits in a few faint red objects, which may turn
out to be difficult to observe with echelle spectrographs. OGLE-TR-82 is
one example, a difficult but interesting case that we have followed up here.

We acquired good seeing images in the $V$-band with VIMOS at the ESO VLT, and 
in the $K$-band with SofI at the ESO NTT.  
We also observed a single transit of this star in the $g$ and $i$-bands with the
GEMINI-S telescope.
Our data is complemented by data of Udalski \etal (2003)
in the $I$-band, who observed 22 transits, but with few points per transit.

We conclude that it was not possible to measure velocities for this
star in the past, and thus to estimate the mass of the companion,
because the star is too faint in the optical ($V=20.6$).

The transit amplitudes are well measured, with  30 points in
transit for each of the two optical bands.  We confirm 
the amplitude of the transit measured by OGLE in the $I$-band, $A_i=0.034 ~$, 
but find that the amplitude in the $g$-band is larger, $A_g=0.10 ~mag$.

We explore different possibilities for this system. 
First, the hypothesis of an M7V primary leads to a possible 
planetary size for the transiting candidate.
In this case, based on the new photometric data, and assuming a M7V primary,
the radius for the
companion would be $R_p=0.3\pm 0.1 ~R_J$, i.e. the size of Neptune.
This scenario has the problem that the measured transit time is too long,
and that the amplitudes seem to be different in the $g$ and $i$-bands.

Second, the
alternative explanation consists of a triple system, composed of an eclipsing binary blended
with a background red giant.
In this case the red giant has to be very reddened 
and very distant.

Clearly, spectroscopic measurements were still needed for this interesting target.
These are more efficiently carried out in the near-infrared, given the
extreme red color of this object. Quantitative analysis of the near-IR spectrum obtained at the ESO NTT finally proved that OGLE-TR-82 is a distant reddened metal poor early giant.  This result is confirmed by direct comparison with stellar templates that gives the best fit with a K3III star with an absorption of $A_V=7 ~mag$.
This rules out a planetary size companion. 

We conclude that this system is a main-sequence binary blended with a background
red giant.

Based on the $I$-band photometry, the OGLE-TR-82 system perfectly
mimics an M dwarf with a Neptune-size companion. After long effort
follow-up, we have found that the system is composed of 
an eclipsing binary blended with a background K giant. There is a lesson to be learned for
future transit surveys searching for hot Neptunes and super Earths around
late type stars.

\acknowledgments
SR, JMF, DM, GP, MZ, MTR, WG are supported by Fondap
Center for Astrophysics No. 15010003. 
We thank the GEMINI staff at Cerro Pachon, the ESO staff at Paranal 
Observatory, and at La Silla Observatory and Ricardo Salinas for his help.

\clearpage

\begin{deluxetable}{lcl}
\tablewidth{0pt}
\tablecaption{Log of observations of OGLE-TR-82}
\tablehead{\colhead{Instrument} & \colhead{Observation Date} & \colhead{}}
\startdata
VLT+VIMOS	&Apr 2005&$V$-band photometry\\
NTT+SofI	&May 2005&$Ks$-band photometry\\
		&         & NIR spectroscopy\\
GEMINI-S+GMOS	&Jan 2006&$g$ \& $i$-band photometry\\
\enddata
\end{deluxetable}

\begin{deluxetable}{lcl}
\tablecaption{Parameters of OGLE-TR-82 and its companion}
\tablehead{\colhead{Parameter} & \colhead{} &\colhead{References}}
\tablewidth{0pt}
\startdata
$Period$&$0.764244 ~d$	&revised by Udalski \etal 2005\\
JDo&$2452323.84747 ~d$  &revised by Udalski \etal 2005\\
$A_I$	&$0.034$	&Udalski \etal 2002\\
$A_i$   &$0.034$        &This work \\
$A_g$   &$ 0.1$         &\\
$V$	&$20.61\pm 0.10$&\\
$I$	&$16.30\pm 0.10$&Udalski \etal 2002\\
$K$	&$12.20\pm 0.10$&\\
$V-I$	&$4.31\pm 0.10$ &\\
$I-K$	&$4.10\pm 0.10$ &\\
$V-K$	&$8.41\pm 0.10$ &\\
$t_T$	&$0.063 ~d$	&\\
\enddata
\end{deluxetable}

\clearpage

\end{document}